\documentclass{iopart}
\usepackage{iopams}
\usepackage{amssymb}
\usepackage{bm}
\usepackage{graphicx}
\begin{document}
\title{Quantum numbers and spectra of structured light}
\author{Iwo Bialynicki-Birula}
\address{Center for Theoretical Physics, Polish Academy of Sciences\\
Aleja Lotnik\'ow 32/46, 02-668 Warsaw, Poland}
\ead{birula@cft.edu.pl}
\author{Zofia Bialynicka-Birula}
\address{Institute of Physics, Polish Academy of Sciences\\
Aleja Lotnik\'ow 32/46, 02-668 Warsaw, Poland}

\begin{abstract}
It is shown that the description of light beams in terms of the corresponding photon quantum numbers elucidates the properties of these beams. In particular, this description shows that the helicity quantum number plays the fundamental role. This mode of description is applied to twisted and knotted electromagnetic waves. We concentrate on the cases where photon wave functions are eigenfunctions of one component of angular momentum. We discovered that for knotted waves the eigenvalue of the angular momentum determines the topology of knots.
\end{abstract}

\vspace{0.5cm}

\hspace*{6cm}Motto\\
\hspace*{6cm}{\em``Quantum optics is that branch of optics\\}
\hspace*{6cm}{\em where the quantum features of light matter''\\}
\hspace*{6.8cm}Wolfgang P. Schleich in his book:\\
\hspace*{6.8cm}Quantum Optics in Phase Space\\

\section{Introduction}

It has been argued in \cite{bb0} that the description of light beams in terms of constituent photons elucidates such properties as helicity, orbital angular momentum and spin; the physical properties of light can be aptly characterized in terms of photon wave functions. The simplest description of photon states is in terms of wave functions in {\em momentum space}. According to Wigner \cite{wig}, spinning massless particles are described by two wave functions $f_\pm(\bm k)$; each of them forms one-dimensional irreducible representation of the proper Poincar\'e group (without reflections). Under reflections, these two wave functions exchange their places. In the present paper we apply this mode of description to structured light; both twisted and knotted.

Twisted light beams are usually characterized by the well-defined component of the total angular momentum in the direction of propagation. In those cases the photon wave functions are eigenfunctions of this component of the angular momentum operator. Such light beams are important in various applications and they are now easily produced with the use of holograms or spatial light modulators (SLM's).

Knotted solutions of Maxwell equations characterized by their stable topological structure have also well defined helicity and one component of the angular momentum. Its eigenvalue determines the topological type of the knots. The knotted waves have not so far been created in experiments. However, the complete Fourier (spectral) analysis determined in this work may be helpful in the construction of spatial light modulators and holograms that will produce knotted electromagnetic waves.

We describe twisted and knotted electromagnetic waves in terms of constituent photons. What distinguishes our analysis from a very thorough discussion of knotted light waves, described in \cite{kl,hss}, is the reliance on the particle aspects of such analysis (especially the photon angular momentum).

\section{Connection between the photon wave function and the electromagnetic field}

Both representations of the Poincar\'e group are needed to describe a general state ${\bm{\mathfrak f}}({\bm k})$ of the photon,
\begin{eqnarray}\label{pwf}
{\bm{\mathfrak f}}({\bm k})=\left(\begin{array}{c}f_+({\bm k})\\f_-({\bm k})\end{array}\right).
\end{eqnarray}
The components of this two-dimensional vector are characterized by the opposite values of helicity: $\pm 1$ in units of $\hbar$. The dimensionless helicity operator ${\hat\lambda}$ for photons acts on ${\bm{\mathfrak f}}({\bm k})$ as follows:
\begin{eqnarray}\label{hel}
{\hat\lambda}\left(\begin{array}{c}f_+({\bm k})\\f_-({\bm k})\end{array}\right)=\left(\begin{array}{c}f_+({\bm k})\\-f_-({\bm k})\end{array}\right).
\end{eqnarray}
Helicity can be considered to be the most fundamental {\em quantum number} for massless particles since it distinguishes between the two {\em inequivalent} representations of the Poincar\'e group.

The electromagnetic field in free space, composed of photons characterized by the pair of wave functions $f_\pm(\bm k)$, can be described by the Riemann-Silberstein (RS) vector ${\bm F}(\bm r,t)$ \cite{bb0,pwf,bb1,bb2},
\begin{eqnarray}
{\bm F}(\bm r,t)&=\frac{{\bm D}}{\sqrt{2\epsilon_0}}+\rmi\frac{{\bm B}}{\sqrt{2\mu_0}}\label{rs}\\
&=\int\!\frac{d^3k}{(2\pi)^{3/2}}\,{\bm e}(\bm k)\left[f_+(\bm k)\rme^{\rmi\bm k\cdot\bm r-\rmi\omega t}+f_-^*(\bm k)\rme^{-\rmi\bm k\cdot\bm r+\rmi\omega t}\right]\label{con},
\end{eqnarray}
where ${\bm e}({\bm k})$ is a complex polarization vector. In order to guarantee that ${\bm F}(\bm r,t)$ obeys the complex version of Maxwell equations (c=1),
\begin{eqnarray}\label{eq}
\rmi\partial_t{\bm F}(\bm r,t)={\bm\nabla}\times{\bm F}(\bm r,t),
\end{eqnarray}
the polarization vector must satisfy the condition:
\begin{eqnarray}\label{eqe}
{\bm k}\times{\bm e}({\bm k})=-\rmi k\,{\bm e}({\bm k}),
\end{eqnarray}
where $k=\omega=\sqrt{k_x^2+k_y^2+k_z^2}$. It is convenient to assume that the polarization vector is normalized, ${\bm e}^*({\bm k})\cdot{\bm e}({\bm k})=1$. This still leaves an undetermined overall phase. In this Section we will use the polarization vector which is the complexified form of the original Whittaker construction \cite{whitt,bb1}:
\begin{eqnarray}\label{e1}
{\bm e}_{\rm W}({\bm k}) = \frac{1}{\sqrt{2}\,\omega\,k_\perp}
\!\left[\begin{array}{c}\vspace{0.2cm}
-k_xk_z+\rmi\omega k_y\\\vspace{0.2cm}
-k_yk_z-\rmi\omega k_x\\
k_\perp^2\end{array}
\right],
\end{eqnarray}
where $k_\perp=\sqrt{k_x^2+k_y^2}$.

The notion of helicity can be extended to the classical electromagnetic field. Namely, the general solution (\ref{con}) of Maxwell equations in free space has two parts: the positive frequency part (associated with the positive helicity wave function $f_+$) and the negative frequency part (associated with the negative helicity wave function $f_-$). However, the separation of the field into two helicity parts without the Fourier representation requires the application of a nonlocal operation \cite{bb3}.

Under the choice (\ref{e1}) of the polarization vector, the Fourier representation (\ref{con}) of the RS vector can be written in the following form:
\begin{eqnarray}\label{rs1}
{\bm F}(\bm r,t)=\left[\begin{array}{c}\vspace{0.2cm}
\partial_x\partial_z+\rmi\partial_y\partial_t\\\vspace{0.2cm}
\partial_y\partial_z-\rmi\partial_x\partial_t\\
-\partial_x^2-\partial_y^2
\end{array}
\right]\chi({\bm r},t),
\end{eqnarray}
where $\chi({\bm r},t)$,
\begin{eqnarray}\label{chi}
\fl\qquad\chi({\bm r},t)=\int\!\frac{d^3k}{(2\pi)^{3/2}}\,
\frac{1}{\sqrt{2}\,\omega\,k_\perp}\left[f_+(\bm k)e^{\rmi\bm k\cdot\bm r-\rmi\omega t}+f_-^*(\bm k)e^{-\rmi\bm k\cdot\bm r+\rmi\omega t}\right],
\end{eqnarray}
is a scalar solution of the d'Alembert equation. The function $\chi({\bm r},t)$ deserves the name of the {\em superpotential} since the electromagnetic field components are obtained as its second derivatives.

The important quantum numbers of the photon wave functions are the eigenvalues of the generators of the Poincar\'e group. Relativistic nature of photons makes the set of quantum numbers associated with the generators of the Poincar\'e group especially significant.

The generators acting on the photon wave functions in the momentum representation have the form (cf., for example, \cite{bb0}):
\numparts
\begin{eqnarray}
{\hat H}=\hbar\omega,\label{gena}\\
{\hat{\bm P}}=\hbar{\bm k},\label{genb}\\
{\hat{\bm J}}=-\rmi\hbar{\bm k}\times{\bm D}+\hbar{\hat\lambda}{\bm n}_{\bm k},\label{genc}\\
{\hat{\bm N}}={\rmi}\hbar\omega\,{\bm D},\label{gend}
\end{eqnarray}
\endnumparts
where ${\bm n}_{\bm k}={\bm k}/k$ and ${\bm D}$ denotes the covariant derivative on the light cone,
\begin{eqnarray}\label{cd}
{\bm D}={\bm{\partial}_{\bm k}}-\rmi{\hat\lambda}{\bm\alpha}({\bm k}).
\end{eqnarray}
The form of ${\bm\alpha}({\bm k})$ depends on the choice of phase of the polarization vector.

A complete specification of the photon state ${\bm{\mathfrak f}}({\bm k})$ requires four quantum numbers: helicity and three eigenvalues of mutually commuting operators. There are only few choices of the generators of the Poincar\'e group that satisfy the condition of commutativity.

\section{Quantum numbers and spectral properties of twisted beams}

The simplest electromagnetic waves are circularly polarized monochromatic plane waves. Their photon wave functions are eigenfunctions of the helicity belonging to the quantum numbers: $\lambda=1$ (for left-handed polarization) or $\lambda=-1$ (for right-handed polarization) and the three components of momentum ${\hat{\bm P}}$ belonging to quantum numbers ${\bm q}=\{q_x,q_y,q_z\}$. Despite their simplicity plane waves are not useful in the analysis of waves with a twist.

The most important quantum number characterizing twisted and knotted waves studied in this paper is the eigenvalue of one component of the angular momentum. The notion of a light beam implies the presence of a distinguished direction; the direction of the beam propagation. This direction is usually chosen along the $z$-axis.

The total angular momentum in momentum representation (\ref{genc}) has two parts: the orbital part perpendicular to $\bm k$ and the helicity part parallel to $\bm k$. We have chosen the phase of the polarization vector (\ref{e1}) in such a way that $J_z$ has the simplest possible form. This choice leads to:
\begin{eqnarray}\label{jz}
J_z=-\rmi\hbar(k_x\partial_{k_y}-k_y\partial_{k_x}).
\end{eqnarray}
It may seem surprising that the helicity component of $J_z$ disappeared. This does not mean that the helicity part in (\ref{genc}) does not play any role. It is only in the formula for the $z$-component of the angular momentum that the ${\bm\alpha}$ part in $\bm D$ cancels the helicity part. This cancelation is due to a particular choice of the phase of the polarization vector (\ref{e1}). We must still remember, however, that the eigenvalues of $J_z$, despite its form (\ref{jz}) that looks like the orbital part, correspond to the $z$-component of the {\em total} angular momentum.

In the description of beams we shall use cylindrical coordinates. The general form of the superpotential $\chi({\bm r},t)$ for a twisted beam with a given eigenvalue $\hbar M$ of the $z$-component of the total angular momentum $J_z$ is:
\begin{eqnarray}\label{cyl}
\chi_{\lambda M}(\rho,\phi,z,t)
&=&\frac{(-\rmi)^M}{2\pi}\!\!\int_0^\infty\!\!\!\! dk_\perp k_\perp\!\int_{-\infty}^{\infty}\!\!\!\!\!\!dk_z\!
\int_0^{2\pi}\!\!\!\!\!d\varphi\nonumber\\
&\times& e^{-\rmi\lambda(\omega t-k_z z-k_\perp\rho\cos(\varphi-\phi))}e^{\rmi M\varphi}g(k_\perp,k_z),
\end{eqnarray}
where $\omega=\sqrt{k_\perp^2+k_z^2}$. We have considered here separately the two helicity parts. Of course, the general solution contains both parts. The function $g(k_\perp,k_z)$ determines directly the spectral composition of the beam. The function $e^{\rmi M\varphi}g(k_\perp,k_z)$ is the photon wave function (up to normalization).

The integration over $\varphi$ in (\ref{cyl}) can be performed and we obtain the following integral representation in terms of Bessel functions,
\begin{eqnarray}\label{cyl1}
\fl\qquad\chi_{\lambda M}(\rho,\phi,z,t)
=e^{\rmi\lambda M\phi}\int_0^\infty\!\!\!\! dk_\perp\!\int_{-\infty}^{\infty}\!\!\!\!dk_z
e^{-\rmi\lambda(\omega t-k_z z)}g(k_\perp,k_z)J_{M}(k_\perp\rho).
\end{eqnarray}
We shall choose this representation of $\chi$ to generate several important beams.

The simplest example of a twisted wave is the Bessel beam. It is obtained from the integral representation (\ref{cyl1}) by choosing the spectral function in the form:
\begin{eqnarray}\label{bes2}
g^{\rm B}(k,k_z)=1/k_\perp\delta(k_\perp-q_\perp)\delta(k_z-q_z).
\end{eqnarray}
Thus, every Bessel beam is monochromatic and is characterized by the following quantum numbers: the helicity $\lambda$, the component of momentum $q_z$ in the direction of propagation, the length of momentum $q_\perp=\sqrt{q_x^2+q_y^2}$ in the perpendicular direction and the $z$-component of the total angular momentum $M$. The superpotential for the Bessel beam has the form:
\begin{eqnarray}\label{bes1}
\chi^{\rm B}_{\lambda Mq_z q_\perp}(\rho,\phi,z,t)
 = e^{-i\lambda(\omega t- q_z z-M\phi)}J_{M}(q_\perp\rho).
\end{eqnarray}

Bessel beams share with plane waves one unrealistic property: they carry infinite energy flux. To overcome this problem, Laguerre-Gauss (LG) beams were introduced to describe more realistic electromagnetic waves. These beams are considered usually only in the paraxial approximation. Here we give a new solution that goes beyond the paraxial approximation. The superpotential $\chi^{\rm LG}$ (for positive helicity) which solves {\em exactly} the d'Alembert equation is:
\begin{eqnarray}\label{lg}
\fl\qquad\chi^{\rm LG}_{qMn}(\rho,\phi,z,t)
=\rme^{\rmi q\left(z-t\right)}\rme^{\rmi M\phi}\rho^{|M|}d_+^{n+|M|+1}d_-^{-n}\rme^{-d_+\rho^2 }
L_n^{|M|}[(d_++d_-)\rho^2],
\end{eqnarray}
where $d_\pm=1/(l^2\pm\rmi(z+t)/q)$. The solution for negative helicity is obtained by complex conjugation. LG beams are characterized by the following quantum numbers: helicity, the $z$-component of the total angular momentum $M$, $n$ and $q$. The characteristic frequency $q$ is the eigenvalue of the operator $\rmi\lambda(\partial_t-\partial_z)/2$. The quantum number $n$ characterizes the shape of the beam in the transverse plane. The parameter $l$ fixes the scale of the Gaussian envelope.

A remarkable property of $\chi^{\rm LG}$ is that at each time $t$ it is also a solution of the paraxial equation and it coincides with the standard expression used in literature on optical beams. In other words, the solution (\ref{lg}) can be obtained from the solution in the paraxial approximation (for example, Eq.~(2.8) of Ref.~\cite{ba}) by the substitution $z\to z+t$ and the insertion of the phase factor $\exp\rmi k(z-t)$.

LG beams are not monochromatic. Their spectral properties in the general case are rather complicated and we give here only the formula for the simplest beam: the pure Gaussian beam obtained for $n=0$. In this case, the spectral function $S(\omega)$ is (disregarding the normalization):
\begin{eqnarray}\label{sp1}
S(\omega) =(\omega-q)^{|M|/2}e^{-l^2q(\omega-q)}.
\end{eqnarray}
The troublesome property of the exact LG beams is that they describe a superposition of beams running in opposite directions.

There exist exact solutions of Maxwell equations describing twisted beams carrying finite energy and not containing the components running in the opposite direction: the exponential beams \cite{bb4}. Their scalar superpotentials for positive helicity are:
\numparts\label{exp}
\begin{eqnarray}
\chi^{\rm Exp1}_{Mq_z}(\rho,\phi,z,t) = \rme^{\rmi(q_z z+M\phi)}\rho^M K_{M+1/2}(|q_z|s)/s^{M+1/2},\\
\chi^{\rm Exp2}_{Mq_z}(\rho,\phi,z,t) =\frac{\rme^{\rmi(q_z z+M\phi)}\rme^{-\vert q_z\vert s}\rho^M}{s\left(s-(\tau+\rmi t)\right)^M},
\end{eqnarray}
\endnumparts
where $s=\sqrt{\rho^2-(t-\rmi\tau)^2}$ and the parameter $\tau$ determines the extension of the beam in the transverse direction. The Macdonald function $K_{M+1/2}/s^{M+1/2}$ is, in fact, an exponential function $\exp(-(|q_z|s))$ multiplied by a polynomial in the inverse powers of $|q_z|s$. The spectral functions for the exponential beams are:
\numparts
\begin{eqnarray}
S_1(\omega)=(\omega^2-(|q_z|)^2)^{M/2}\rme^{-\omega\tau},\label{exp1}\\
S_2(\omega)=\left(\frac{\omega-|q_z|}
{\omega+|q_z|}\right)^{M/2}\rme^{-\omega\tau}.\label{exp2}
\end{eqnarray}
\endnumparts
Like Bessel beams, the exponential beams have quantum numbers $\lambda$, $M$ and $q_z$ but, of course, they are not monochromatic. An interesting feature of the spectral functions (\ref{sp1}), (\ref{exp1}) and(\ref{exp2}) is the presence of the threshold $\omega>q_z$.

The scalar superpotentials for the opposite values of helicity in our three examples differ only by complex conjugation. This is the general property and it follows directly from the formula (\ref{chi}).

\section{Quantum numbers and spectral properties of knotted waves}\label{knots}

Eigenfunctions of one component of the angular momentum play a role not only for collimated beams. Also some localized solutions of Maxwell equations with intricate topological structure have similar properties. A large class of such solutions was found in \cite{kl} but here we shall only restrict ourselves to the simplest cases.

In 1956 in his book \cite{syn} Synge described a localized solution of Maxwell equations which he tried to interpret as a model of an electron. Many years later \cite{ran} Ra{\~n}ada rediscovered this solution (now called Hopfion) and found its topological properties connecting it with the Hopf fibration. In all examples of knotted EM waves considered here the Hopfion will be used as a fundamental building block.

Hopfions for both helicities $\lambda$ can be derived from the superpotentials $\chi_{\rm H}(\bm r,t)$,
\begin{eqnarray}\label{hop}
\fl\qquad\chi_{\rm H}(\bm r,t)=\frac{1}{x^2 + y^2 + z^2-(t-\rmi a)^2}=\frac{1}{4\pi}\int\!\frac{d^3k}{k}\,
e^{-|a|k}\rme^{\rmi\lambda(\bm k\cdot\bm r-\omega t)},
\end{eqnarray}
where $\lambda={\rm sgn}(a)$ and this means that the sign of $a$ determines the sign of helicity of the solutions obtained from these superpotentials.

Various configurations of the electromagnetic field can be obtained from $\chi_{\rm H}$
depending on the choice of the polarization vector. We found it most convenient for the knotted beams to use the normalized polarization vector in the form:
\begin{eqnarray}\label{e2}
{\bm e}_{\rm H}({\bm k}) = \frac{1}{2\sqrt{2}\,k\,l_+}
\!\left[\begin{array}{c}\vspace{0.2cm}
k_+^2-l_+^2\\\vspace{0.2cm}
-\rmi(k_+^2+l_+^2)\\
2k_+ l_+\end{array}
\right],
\end{eqnarray}
where $k_\pm=k_x\pm\rmi k_y$ and $l_\pm=k\pm k_z$. The polarization vector ${\bm e}_{\rm H}$ differs from the polarization vector ${\bm e}_{\rm W}$ by the phase factor, ${\bm e}_{\rm H}=(k_x+\rmi k_y)/k_\perp{\bm e}_{\rm W}$. The RS vector constructed from the superpotential (\ref{hop}) is:
\begin{eqnarray}\label{rs2}
{\bm F}_{\rm H}(\bm r,t)=\frac{1}{d^3}\left[\begin{array}{c}\vspace{0.2cm}
t_+^2-x_+^2\\
\vspace{0.2cm}
\rmi(t_+^2+x_+^2)\\
-2t_+x_+
\end{array}
\right],
\end{eqnarray}
where $t_\pm=t\pm z-\rmi a$, $x_\pm=x\pm\rmi y$ and
\begin{eqnarray}\label{s}
d=x^2 + y^2 + z^2-(t-\rmi a)^2=x_+x_--t_+t_-.
\end{eqnarray}
The Fourier representation of the Hopfion RS vector is:
\begin{eqnarray}\label{pm}
\fl\qquad{\bm F}_{\rm H}(\bm r,t)=\frac{1}{4\pi}\int\!\frac{d^3k}{k}
\left[\begin{array}{c}\vspace{0.2cm}
k_+^2-l_+^2\\ \vspace{0.2cm}
-\rmi(k_+^2+l_+^2)\\
2k_+l_+
\end{array}\right]e^{-|a|k}
\rme^{\rmi\lambda(k_+x_-+k_-x_+-l_+t_--l_-t_+)/2},
\end{eqnarray}
The photon wave function $e^{-|a|k}/k$ in (\ref{hop}) is spherically symmetric, however, one cannot argue that  $J_z$ vanishes because this wave function is now multiplied by a different polarization vector. It turns out that for this choice $J_z$ has also the helicity part,
\begin{eqnarray}\label{jz1}
J_z=-\rmi\hbar(k_x\partial_{k_y}-k_y\partial_{k_x})+\hbar\lambda,
\end{eqnarray}
and helicity contributes one unit to the total angular momentum. Thus, the Hopfion carries angular momentum $\hbar\lambda$ along the $z$-directions. The observation that the Hopfian solution carries angular momentum has been made before (cf., for example, \cite{ib,hss}) but only with the use of the total angular momentum defined as an integrated density ${\bm r}\times({\bm E}\times{\bm B})$.

The RS vector (\ref{rs2}) represents a null electromagnetic field, i.e.:
\begin{eqnarray}\label{null}
{\bm F}_{\rm H}^2=\epsilon\left(({\bm E}^2-{\bm B^2})/2+\rmi {\bm E}\cdot {\bm B}\right)=0.
\end{eqnarray}
The previous choice (\ref{e1}) of the polarization vector would not lead to a null field, whereas null fields are essential for topological stability of knots.

\begin{figure}[t]
\centering
\includegraphics[width=12cm,height=2.4cm]{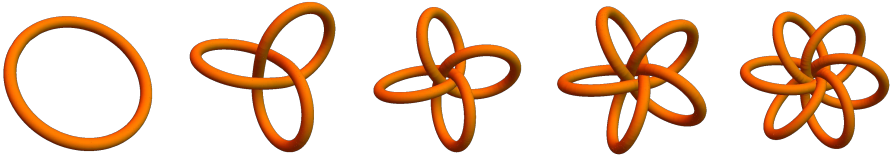}
\caption{The knotted lines of the electric field at $t=0$\\for the following choices of the coefficients $n_1$ and $n_2$\\in Eq.~(\ref{hop1}):\,(1,0),\,(1,5),
\,(1,7),\,(1,4),\,and\,(1,11).}
\label{fig1}
\end{figure}

Single Hopfions do not give field lines of interesting topological structure, but by superposition of Hopfions  we may obtain waves whose field lines exhibit all kinds of knots. For example, the superposition $\bm F_{(n_1,n_2)}$ of two ${\bm F}_{\rm H}$ vectors with positive and negative values of helicity ($a=\pm 1$) exhibits such structures,
\begin{eqnarray}\label{hop1}
{\bm F}_{(n_1,n_2)}(\bm r,t)=n_1{\bm F}_{\rm H}^{(a=1)}+n_2{\bm F}_{\rm H}^{(a=-1)}.
\end{eqnarray}
In Fig.~\ref{fig1} we plotted the lines of the electric field for various choices of the coefficients. Unfortunately, those knots become untied during the time evolution in agreement with the common wisdom that stable knots exist only for null electromagnetic fields.

Topologically stable knots were found in \cite{kl} with the use of the Bateman construction \cite{bat} of null electromagnetic fields. These solutions are:
\begin{eqnarray}\label{kl}
{\bm F}_{pq}(\bm r,t)={\bm\nabla}\alpha^p\times{\bm\nabla}\beta^q,
\end{eqnarray}
where $(p,q)$ are coprime numbers. The Bateman functions $\alpha$ and $\beta$ were chosen in \cite{kl} as:
\begin{eqnarray}\label{def}
\alpha=1-2\rmi at_-/d,\qquad\beta=2ax_-/d.
\end{eqnarray}
To avoid negative values of the angular momentum, we shall rotate this solution by 180 degrees around the $x$-axis to obtain:
\begin{eqnarray}\label{def1}
\alpha=1-2\rmi at_+/d,\qquad\beta=2ax_+/d.
\end{eqnarray}
In order to find the photon wave functions of the knotted solutions, we rewrite ${\bm F}_{pq}(\bm r,t)$ in the form:
\begin{eqnarray}\label{kl1}
{\bm F}_{pq}(\bm r,t)=pq\alpha^{p-1}\beta^{q-1}
{\bm\nabla}\alpha\times{\bm\nabla}\beta
=pq\alpha^{p-1}\beta^{q-1}{\bm F}_{\rm H}(\bm r,t).
\end{eqnarray}
The electromagnetic field described by this vector is null because a null vector ${\bm F}_{\rm H}$ is multiplied by a scalar function.

From the representation (\ref{kl1}) of knotted waves, by a straightforward procedure, we can obtain the photon wave functions. Our derivation is based on the observation that the factors $\alpha$ and $\beta$ multiplying the RS vector ${\bm F}_{\rm H}(\bm r,t)$ can be expressed in terms of the partial derivatives $\partial/\partial x_-$ and $\partial/\partial t_-$, respectively. To this end we rewrite the factor $\alpha^{p-1}\beta^{q-1}$ in the form:
\begin{eqnarray}\label{fac}
\alpha^{p-1}\beta^{q-1}
=\sum_{r=0}^{p-1}(-1)^r\frac{(p-1)!}{r!(p-r-1)!}
\gamma^r\beta^{q-1},
\end{eqnarray}
where $\gamma=1-\alpha=2\rmi at_+/d$. Next, we replace each product $\gamma^r\beta^{q-1}{\bm F}_{\rm H}$ by the derivatives acting on ${\bm F}_{\rm H}$,
\begin{eqnarray}\label{der}
\gamma^r\beta^{q-1}{\bm F}_{\rm H}=\frac{2(-1)^{q-1}}{(q+r+1)!}\left(2\rmi a\frac{\partial}{\partial t_-}\right)^r\left(2a\frac{\partial}{\partial x_-}\right)^{q-1}{\bm F}_{\rm H}.
\end{eqnarray}

The derivatives in this formula are easily evaluated using the Fourier representation (\ref{pm}). Finally, the RS vector for knotted waves takes the form (we assumed positive helicity):
\begin{eqnarray}\label{pq1}
\fl\qquad{\bm F}_{pq}(\bm r,t)=\!\int\!\frac{d^3k}{k}
\!\left[\begin{array}{c}\vspace{0.2cm}
k_+^2-l_+^2\\ \vspace{0.2cm}
-\rmi(k_+^2+l_+^2)\\
2k_+l_+
\end{array}\right]\!
\rme^{\rmi(q-1)\varphi}k_\perp^{q-1}g_{pq}(l_+)
\rme^{-|a|k}\,\rme^{\rmi(\bm k\cdot\bm r-\omega t)},
\end{eqnarray}
where $g_{pq}(l_+)$ is a polynomial of the degree $p-1$.
The presence of the phase factor $\rme^{\rmi(q-1)\varphi}=[(k_x+\rmi k_y)/k_\perp]^{(q-1)}$ means that the orbital angular momentum in the $z$-direction is $(q-1)\hbar$. Therefore, the total angular momentum in the $z$-direction is $\hbar M=\hbar q$ (one additional unit is contributed by the helicity of the Hopfian).

The three examples of knotted waves described in \cite{kl}: the trefoil $(p=2,q=3)$, the cinquefoil $(p=2,q=5)$ and four linked rings $(p=2,q=2)$ have different angular momentum quantum numbers $M=3, M=5, M=2$, respectively.

The spectrum of knotted waves still contains the exponential factor $\rme^{-|a|k}$ but is more complicated than for the pure Hopfian because the photon wave function in this case depends not only on $\omega$ but also on $k_\perp$ and $l_+$.

\section{Conclusions}

The direct correspondence between the Fourier representation of the electromagnetic field and the photon wave functions enables us to define precisely the helicity of the classical field. The use of the photon quantum numbers offers a classification scheme for twisted and knotted optical waves. In particular, in the examples considered here the photon wave functions were eigenstates of the $z$-component of the total angular momentum. The eigenvalue of the $z$-component of the angular momentum determines the topological properties of knotted waves. Moreover, in all cases studied here, the photon wave functions in momentum representation are much simpler than the corresponding fields in space-time.

\end{document}